\documentclass[aps,prc,letterpaper,11pt
,twoside,tightenlines,nofootinbib,showpacs
,preprint]{revtex4-1}
\usepackage{graphicx}
\usepackage{color}
\usepackage[sort&compress]{natbib}
\usepackage{subfigure}
\usepackage{amsmath}
\usepackage{amssymb}
\usepackage{amsfonts}
\usepackage{cancel}
\usepackage{hyperref}
\usepackage{fontenc}
\begin{document}

\arraycolsep1.5pt

\newcommand{\Ima}{\textrm{Im}}
\newcommand{\Rea}{\textrm{Re}}
\newcommand{\mev}{\textrm{ MeV}}
\newcommand{\be}{\begin{equation}}
\newcommand{\ee}{\end{equation}}
\newcommand{\ba}{\begin{eqnarray}}
\newcommand{\ea}{\end{eqnarray}}
\newcommand{\gev}{\textrm{ GeV}}
\newcommand{\nn}{{\nonumber}}
\newcommand{\dtres}{d^{\hspace{0.1mm} 3}\hspace{-0.5mm}}

\def\del{\partial}

\title{Meson baryon components in the states of the baryon decuplet}

\author{F. Aceti,$^{1,3}$ L. R. Dai,$^{2,3}$ L. S. Geng,$^4$ E. Oset$^{1,3}$ and Y. Zhang$^{2}$}
\affiliation{
$^{1}$Departamento de F\'{\i}sica Te\'orica and IFIC, Centro Mixto Universidad de 
Valencia-CSIC,
Institutos de Investigaci\'on de Paterna, Apartado 22085, 46071 Valencia,
Spain\\
\\
$^{2}$Department of Physics, Liaoning Normal University, Dalian 116029, China\\
\\
$^{3}$Kavli Institute for Theoretical Physics China, CAS, Beijing 100190, China\\
\\
$^{4}$School of Physics and Nuclear Energy Engineering and International Research Center for Nuclei and Particles in the Cosmos, Beihang University, Beijing 100191, China }

\date{\today}

\begin{abstract}
We apply an extension of the Weinberg compositeness condition on partial waves of $L=1$ and resonant states to determine the weight of meson-baryon component in the $\Delta(1232)$ resonance and the other members of the $J^P= \frac{3}{2}^+$ baryon decuplet. We obtain an appreciable weight of $\pi N$ in the $\Delta(1232)$ wave function, of the order of 60 \%, which looks more natural when one recalls that experiments on deep inelastic and Drell Yan give a fraction of $\pi N$ component of 34 \% for the nucleon. We also show that, as we go to higher energies in the members of the decuplet, the weights of meson-baryon component decrease and they already show a dominant part for a genuine, non meson-baryon, component in the wave function. We write a section to interpret the meaning of the Weinberg sum-rule when it is extended to complex energies and another one for the case of an energy dependent potential.

\end{abstract}

\maketitle

\section{Introduction}
\label{Intro}

The investigation of the structure of the different hadronic states is one of the most important topics in hadron spectroscopy.
In order to describe the rich spectrum of excited hadrons quoted in the PDG \cite{pdg}, the traditional concept that mesons and baryons are composed, respectively, of two or three quarks, has been replaced by more complex interpretations, like the ones involving more quarks \cite{more1,more2}.

A remarkable success in describing hadron structure has been obtained by chiral perturbation theory ($\chi PT$) \cite{weinchiral, chiral}, an effective field theory in which the building blocks are the ground state mesons and baryons. The low energy processes are well described in this framework, but its limited energy range of convergence makes it unsuitable to deal with higher energies.

Combining unitarity constraints in coupled channels of mesons and baryons with the use of chiral Lagrangians, an extension of the theory to higher energies was made possible. The resulting theory, usually referred to as chiral unitary approach \cite{Kaiser:1995eg,Kaiser:1996js,npa,Kaiser:1998fi,ramonet,angelskaon,ollerulf,Jido:2002yz,carmen,cola,carmenjuan,hyodo, review}, allows to explain many mesons and baryons in terms of the meson-meson and meson-baryon interactions provided by chiral Lagrangians, interpreting them as composite states of hadrons. This kind of resonances are commonly known as ``dynamically generated".

An interesting challenge in the study of the hadron spectrum, is understanding whether a resonance can be considered as a composite state of other hadrons or else a ``genuine" state. An early attempt to answer this question was made by Weinberg in a time honored work \cite{weinberg}, in which it was determined that the deuteron was a composite state of a proton and a neutron. Other works on this issue are \cite{han1, han2, cleven}. However, the analysis was made in the case of $s$-waves and for small binding energies. An extension to larger binding energies, using also coupled channels and in the case of bound states, was done in \cite{gamermann}, while in \cite{yamagata} also resonances are considered. 

In a recent paper, the work was generalized to higher partial waves \cite{aceti} and the results obtained were used to justify the commonly accepted idea that the $\rho$ meson is not a $\pi\pi$ composite state but a genuine one. The same method was also successfully used in \cite{xiao} to evaluate the weight of composite $K\pi$ state in the $K^{*}$ wave function. However, no attempt was done to apply the method to baryonic resonances. We use it here to investigate the nature of the baryons of the $J^{P}=\frac{3}{2}^{+}$ decuplet.  

The paper proceeds as follows. In Section \ref{form} we make a brief summary of the formalism. In Section \ref{delta} we address the problem of $\pi N$ scattering in the $\Delta(1232)$ region. In Section \ref{decuplet} we extend the test to all the particles of the decuplet while Section \ref{five} is devoted to discussing and interpreting the meaning of the Weinberg sum-rule when extended to complex energies. Finally, we make some conclusions in Section \ref{concl}.


\section{Overview of the formalism}
\label{form}
The creation of a resonance from the interaction of many channels at a certain energy, takes place from the collision of two particles in a channel which is open. 

The process is described by the set of coupled Schr\"{o}dinger equations,
\begin{equation}
\label{eq:ls_res}
\begin{split}
|\Psi\rangle&=|\Phi\rangle +\frac{1}{E-H_{0}}V|\Psi\rangle\\&=|\Phi\rangle +\frac{1}{E-M_i-\frac{\vec{p}\,^2}{2\mu_i}}V|\Psi\rangle\ ,
\end{split}
\end{equation}
where
\begin{equation}
\label{eq:2}
|\Psi\rangle=
\begin{Bmatrix}
|\Psi_{1}\rangle \\ |\Psi_{2}\rangle \\\vdots \\|\Psi_{N}\rangle
\end{Bmatrix}\ ,\ \ \ \ \ \ \
|\Phi\rangle=
\begin{Bmatrix}
|\Phi_{1}\rangle \\0\\\vdots \\0
\end{Bmatrix}\ ,
\end{equation}
$H_0$ is the free Hamiltonian and $\mu_{i}$ is the reduced mass of the system of total mass $M_{i}=m_{1i}+m_{2i}$. The state $|\Phi_1\rangle$ is an asymptotic scattering state which is used to create a resonance which will decay into other channels.

Since we shall use this in the discussion later on, it is worth stressing that the wave function is defined up to a global phase, the same for $|\Psi\rangle$ and $|\Phi\rangle$, as one can see in Eq.\eqref{eq:ls_res}. However,  the standard prescription is to take for $\Phi$ the $e^{i\vec{k}\vec{r}}$ plane wave function, which then determines the phase of $\Psi$. We shall come back to the question of phases when we use wave functions in the following.

Following \cite{aceti}, we take as the potential $V$
\begin{equation}
\label{eq:pot}
\langle\vec{p}|V|\vec{p}\ '\rangle\equiv (2l+1)\ v\ \Theta(\Lambda-p)\Theta(\Lambda-p')|\vec{p}\,|^{l}|\vec{p}\ '|^{l}P_{l}(\cos\theta)\ ,
\end{equation}
where $\Lambda$ is a cutoff in the momentum space and $v$ is a $N\times N$ matrix, with $N$ the number of channels. The form of the potential is such that the generic $l$-wave character of the process is contained in the two factors $|\vec{p}\,|^{l}$ and $|\vec{p}\ '|^{l}$, and in the Legendre polynomial $P_{l}(\cos\theta)$, so that $v$ can be considered as a constant matrix.

The $N\times N$ scattering matrix, such that $T\Phi=V\Psi$, can be written as
\begin{equation}
\label{eq:tt}
T=(2l+1)P_{l}(\hat{p},\hat{p}')\Theta(\Lambda-p)\Theta(\Lambda-p')|\vec{p}\,|^{l}|\vec{p}\ '|^{l}t\ ,
\end{equation}
and the Schr\"odinger equation leads to the Lippmann-Schwinger equation for $T$ ($T=V+VGT$), by means of which one obtains
\begin{equation}
\label{eq:t}
t=\frac{v}{(1-vG)}=\frac{1}{v^{-1}-G}\ . 
\end{equation}
The matrix $G$ in Eq. \eqref{eq:t} is the loop function diagonal matrix for the two hadrons in the intermediate state (see Eq. \eqref{eq:loop}). Note that the definition  $T\Phi=V\Psi$ makes $T$ independent of the phase convention on the wave function.

The derivation in \cite{aceti} leads to a $t$ matrix which does not contain the factor $|\vec{p}\,|^{l}$, since now the potential $v$ is a constant. Differently from other approaches for $p$-waves, like the one of \cite{ollerpalomar,doring}, which factorize on shell $|\vec{p}\,|^{l}$ and associate it to the potential $v$, this factor is now absorbed in a new loop function
\begin{equation}
\label{eq:loop}
G_{ii}=\int_{_{|\vec{p}\,|<\Lambda}}{d^3p\,\frac{|\vec{p}\ |^{2l}}{E-m_{1i}-m_{2i}-\frac{\vec{p}\,^{2}}{2\mu_i}}}\ ,
\end{equation}
which  is different from the one normally used in the chiral unitary approach \cite{oller}.

This choice is necessary for the generalization of the sum-rule for the couplings, found in \cite{gamermann} for the case of $s$-waves, to any partial wave. Indeed, as shown in \cite{aceti}, for a resonance or bound state dynamically generated by the interaction in coupled channels of two hadrons, the following relationship holds (see an alternative derivation in \cite{jidohyodo})
\begin{equation}
\label{eq:sumrule}
\sum_{i}g_{i}^{2}\left[\frac{dG_{i}}{dE}\right]_{E=E_{R}}=-1\ ,
\end{equation}  
where $E_R$ is the position of the complex pole representing the resonance and $g_i$ is the coupling to the channel $i$ defined as
\begin{equation}
\label{eq:coupling}
g_ig_j=\lim_{E\rightarrow E_R}(E-E_R)t_{ij}\ .
\end{equation}

Note that this definition leads to complex couplings and the sum rule that we derive is 
obtained in terms of them.  

In Section \ref{five} we shall rewrite Eq. \eqref{eq:sumrule} for complex energies and discuss the meaning of each term. We anticipate here that each term represents the integral of the wave function squared  (not the modulus squared) of each component, but this occurs only in a certain phase convention for the wave function that we shall then discuss. The terms of Eq. \eqref{eq:sumrule} are complex, which means that the imaginary parts cancel and then one has 
\begin{equation}
\label{eq:srreal}
\sum_{i}Re\left(g_{i}^{2}\left[\frac{dG_{i}}{dE}\right]_{E=E_{R}}\right)=-1\ ,
\end{equation}
and knowing the meaning of these terms, we can consider each one of them as a measure of the relevance or the weight of a channel in the wave function of the state, but not a probability, which for open channels is not a useful concept since it will diverge.

Sometimes, our knowledge of all needed coupled channels will be incomplete and we shall only have information on hadron-hadron scattering. There can be a genuine component  different to the hadron-hadron one that we study. In order to take into account the weight of this genuine component, Eq. \eqref{eq:sumrule} can be rewritten as
\begin{equation}
\label{eq:sumrule2}
-\sum_{i}Re\left(g_{i}^{2}\left[\frac{dG_{i}}{dE}\right]_{E=E_{R}}\right)=1-Z\ ,\ \ \ \ \ \ \ \ \ \ \ \ Z=Re\int d^3p\left(\Psi_{\beta}(p)\right)^2\ ,
\end{equation}
where $\Psi_{\beta}(p)$ is the genuine component in the wave function of the state, when it is omitted from the coupled channels.

Note that the fixing of a phase in the wave function of one channel will determine the phase of the other wave functions in a coupled set of Lippman-Schwinger equations (see Eqs. \eqref{eq:ls_res} and \eqref{eq:2}). 

The left-hand side of Eq. \eqref{eq:sumrule2} is the measure of this weight of hadron-hadron component, while its diversion from unity measures the weight of something different in the wave function.

The interpretation of $Z$ as a probability for the non meson-baryon component is rigorous for bound states. For poles in the complex plane we have to reinterpret these numbers, as we have mentioned and will be amply discussed in Section \ref{five}.

\section{$\pi N$ scattering and the $\Delta(1232)$ resonance}
\label{delta}
As already mentioned in the Introduction, the sum-rule of Eq. \eqref{eq:sumrule2} has been successfully applied to the $\rho$ and $K^*$ mesons in \cite{aceti} and \cite{xiao}, respectively. We use it for the first time to investigate the nature of a baryonic resonance, the $\Delta(1232)$, in order to quantify the weight of $\pi N$ in this state. 

We first use a model based on chiral unitary theory, and then, we perform a phenomenological test which makes use only of $\pi N$ scattering data. 


\subsection{The model dependent test}
\label{model}
Following the approach of \cite{aceti, xiao} we use a potential of the type
\begin{equation}
\label{eq:pot1}
v=-\alpha\left(1+\frac{\beta}{s_{R}-s}\right)\ ,
\end{equation}
where $\sqrt{s_{R}}$ is the bare mass of the $\Delta$ resonance and $\alpha$ and $\beta$ are two constant factors. Note that we are putting explicitly a CDD pole in $v$ in order to accommodate a possible genuine component of the $\Delta(1232)$ in its wave function \cite{castillejo}. In order to account for the $p$-wave character of the process, the potential $v$ is not dependent on the momenta of the particles. 

Now, we fit the $\pi N$ data for the phase shifts using
\begin{equation}
t=\frac{1}{v^{-1}-G}\ .
\label{eq:tmodel}
\end{equation}
Since the pion is relativistic in the decay of the $\Delta(1232)$, we generalize the equations as already done for the case of $\rho\rightarrow\pi\pi$ in \cite{aceti}.
We take only the positive energy part of the relativistic generalization of the loop function, modified to contain the $|\vec{q}\,|^2$ factor (see Eq. \eqref{eq:loop} and \cite{aceti} for more details),
\begin{equation}
G(s)=\int{\frac{d^3q}{(2\pi)^3}\frac{1}{2\omega(q)}\frac{M_N}{E_N(q)}\frac{q^2}{\sqrt{s}-\omega(q)-E_N(q)+i\epsilon}}\ ,
\label{eq:G}
\end{equation}
with $M_N$ the mass of the nucleon, $m_{\pi}$ the mass of the pion, $E_N(q)=\sqrt{q^2+M_N^2}$ and $\omega(q)=\sqrt{q^2+m_{\pi}^2}$. The loop function in Eq. \eqref{eq:G} is regularized by the cutoff $\theta(\Lambda-|\vec{q}\,|)$ of the potential (see Eq. \eqref{eq:pot}), hence $\Lambda$ plays the role of  $q_{max}$ in the integral of Eq. \eqref{eq:G}.

To be more in agreement with a propagator which has a denominator linear in the energy, we slightly modify Eq. \eqref{eq:pot1} as
\begin{equation}
\label{eq:pot2}
v=-\frac{\alpha}{M_\Delta^4}\left(1+\frac{\beta}{\sqrt{s_{R}}-\sqrt{s}}\right)\ ,
\end{equation}
where the factor $1/M_\Delta^4$ is introduced in order to have both parameters, $\alpha$ and $\beta$, in units of $MeV$. 

The $\pi N$ phase shift is given by the formula \cite{xie}
\begin{equation}
T=p^2t=\frac{-4\pi\sqrt{s}}{M_{N}}\frac{1}{p\cot\delta(p)-ip}\ ,
\label{eq:phaseshift}
\end{equation}
with $p$ the momentum of the particles in the center of mass reference frame.

\begin{figure}[h!]
\includegraphics[width=15cm,height=11cm]{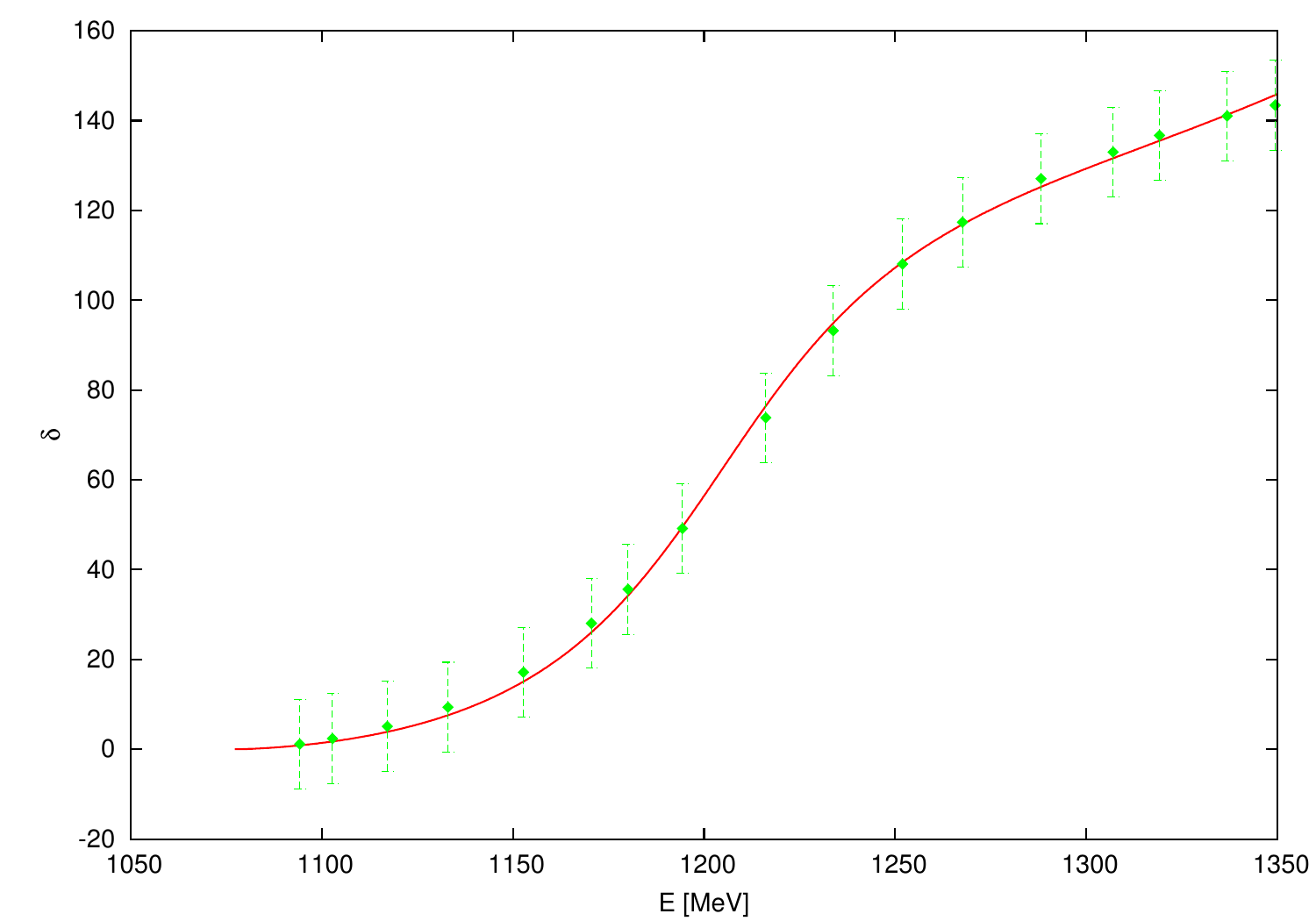}
\caption{The solid curve represents the $\pi N$ scattering $p$-wave phase shifts obtained with the new approach. The data are taken from \cite{cns}.}
\label{fig:phase}
\end{figure}

From the best fit to the $\pi N$ data we obtain the following values of the four parameters:
\begin{equation}
\begin{split}
\label{eq:bfit}
&\alpha=698.0\cdot 10^3\ MeV\ ,\ \ \ \ \ \ \ \ \beta=112.5\ MeV\ ,   \\
&\sqrt{s_R}=1313.8\ MeV\ ,\ \ \ \ \ \ \ \ q_{max}=452.6\ MeV\ .
\end{split}
\end{equation} 
The results of the fit are shown in Fig. \ref{fig:phase}.

Now we want to apply the sum-rule of Eq. \eqref{eq:sumrule2} to our case. We need to extrapolate the amplitude to the complex plane and look for the complex pole $\sqrt{s_0}$ in the second Riemann sheet. This is done by changing $G$ to $G^{II}$ in Eq. \eqref{eq:tmodel}, to obtain $t^{II}$. The function $G^{II}$ is the analytic continuation of the loop function in the second Riemann sheet and is defined as
\begin{equation}
\label{eq:GII}
G^{II}(s)=G^{I}+\frac{i}{2\pi}\frac{M_N}{\sqrt{s}}p^3\ ,\ \ \ \ \ \ Im(p)>0\ ,
\end{equation}
with $G^{I}$ and $G^{II}$ the loop functions in the first and second Riemann sheet, and $G^{I}$ given by Eq. \eqref{eq:G}.

We are now able to obtain the coupling $\tilde{g}_{\Delta}$ as the residue in the pole of the amplitude
\begin{equation}
\label{eq:res}
\tilde{g}_{\Delta}^2=\lim_{\sqrt{s}\rightarrow \sqrt{s_0}}(\sqrt{s}-\sqrt{s_0})t^{II}\ ,
\end{equation}
and to apply the sum-rule of Eq. \eqref{eq:sumrule2} to evaluate the $\pi N$ contribution to the $\Delta$ resonance
\begin{equation}
\label{eq:sr}
-Re\left[\tilde{g}_{\Delta}^2\left[\frac{dG^{II}(s)}{d\sqrt{s}}\right]_{\sqrt{s}=\sqrt{s_0}}\right]=1-Z\ ,
\end{equation}
with $Z$ the weight of something different from a $\pi N$ state in the $\Delta$.

The value of the pole that we get for the best fit is
\begin{equation}
\label{eq:polebf}
\sqrt{s_0}=(1204.6+i44.37)\ MeV\ ,
\end{equation}
while for the coupling we find
\begin{equation}
\label{eq:coupbf}
\tilde{g}_{\Delta}=(8.53+i1.85)\cdot 10^{-3}\ MeV^{-1}\ .
\end{equation}
From these values we finally obtain
\begin{equation}
\label{eq:rulebf}
-\tilde{g}_{\Delta}^2\left[\frac{dG^{II}(s)}{d\sqrt{s}}\right]_{\sqrt{s}=\sqrt{s_0}}=(0.62-i0.41)\ ,
\end{equation}
and
\begin{equation}
\label{eq:modbf}
1-Z=0.62\ ,
\end{equation}
which indicates a sizeable weight of $\pi N$ in the resonance.

\subsection{The phenomenological test}
\label{phen}
Now we want to evaluate the same quantity using a more phenomenological approach. We repeat the analysis of \cite{aceti, xiao} to test the sum-rule by means only of the experimental data.

The $\Delta$ amplitude in a relativistic form is given by
\begin{equation}
\label{eq:phenampl}
t_{\Delta}=\frac{g_{\Delta}^2}{\sqrt{s}-M_{\Delta}+i\frac{\Gamma_{on}}{2}\left(\frac{p}{p_{on}}\right)^3}\ ,
\end{equation}
where
\begin{equation}
p=\frac{\lambda^{1/2}(s,M^2_N, m_{\pi}^2)}{2\sqrt{s}}
\end{equation}
is the three-momentum of the particles in the center of mass reference frame,
\begin{equation}
p_{on}=p(\sqrt{s}=M_{\Delta}),
\end{equation}
and
\begin{equation}
\label{eq:cophen}
g_{\Delta}^2=\frac{2\pi M_{\Delta}\Gamma_{on}}{p_{on}^3 M_{N}}\ .
\end{equation}
The values of $M_{\Delta}$ and $\Gamma_{on}$ are known from the experiment.

Defining $\sqrt{s}=a+ib$ and making the substitution $p\rightarrow -p$ in the width term, we obtain the amplitude $t_{\Delta}$ in the second Riemann sheet. Then, we proceed as before to get the pole and the coupling. 

The values we obtain for the pole and the coupling,
\begin{equation}
\begin{split}
&\sqrt{s_0}=(1208.00+i40.91)\ MeV\ ,\\
&g_{\Delta}=(7.78+i1.86)\cdot 10^{-3}\ MeV^{-1}\ ,
\label{eq:polecoup}
\end{split}
\end{equation}
are very similar to those obtained with  the procedure of the former subsection.

In this case we do not know the size of the cutoff $q_{max}$ needed to regularize the loop function, but the derivative of $G^{II}$ in Eq. \eqref{eq:sr} is logarithmically divergent in the case of $p$-waves. Then, using natural values for the cutoff, as done in \cite{aceti, xiao}, we can establish the stability of the results in a certain range of $q_{max}$.
\begin{table}[ tp ]%
\begin{tabular}{c|c|c}
\hline %
$q_{max}\ [GeV]$ &  $-g^2\frac{dG^{II}}{d\sqrt{s}}$ & $1-Z$\\\toprule %
$0.4$ & $0.47-i0.38$ & $0.47$ \\
$0.5$ & $0.57-i0.29$ & $0.57$ \\
$0.6$ & $0.65-i0.22$ & $0.65$ \\\hline
\end{tabular}
\caption{Values of $-g^2\frac{dG^{II}}{d\sqrt{s}}$ and  $1-Z$ for different cutoffs $q_{max}$.}
\label{tab:z}\centering %
\end{table}

The values of $1-Z$ for three different values of $q_{max}$ are shown in Table \ref{tab:z}. They are rather stable and consistent with the result obtained in the previous section.

\section{Application to other resonances}
\label{decuplet}
Now we extend the study of the hadron-hadron content of resonances to the whole $J^{P}=\frac{3}{2}^{+}$ baryons decuplet. 

We proceed as in the case of the $\Delta(1232)$, applying the phenomenological test of Sec. \ref{phen} to the other particles of the decuplet, $\Sigma(1385)$, $\Xi(1535)$ and $\Omega^{-}$. 

We first investigate the $\pi\Lambda$ and $\pi\Sigma$ content of the $\Sigma(1385)$ wave function. It is known from the PDG \cite{pdg} that it couples to these two channels with different branching ratios: $87\%$ and $11.7\%$ , respectively. In order to evaluate the couplings of the resonance to the single channel, the branching ratios must be taken into account, modifying Eq. \eqref{eq:cophen} as follows:
\begin{equation}
\label{eq:sigcoup}
g_{\Sigma^{*},i}^2=\frac{2\pi M_{\Sigma^{*}}\Gamma_{on}}{p^3_{(i)on} M_{i}}\cdot BR^{(i)}\ , 
\end{equation}
where $BR^{(i)}$ is the branching ratio to the channel $i$, with $i=\pi\Lambda, \pi\Sigma$ and
\begin{equation}
p_{on}^{(i)}=p^{(i)}(\sqrt{s}=M_{\Sigma^*})\ ,
\end{equation}
where
\begin{equation}
p^{(i)}=\frac{\lambda^{1/2}(s,M^2_i, m_{\pi}^2)}{2\sqrt{s}}\ .
\end{equation}

On the other hand, the case of the $\Xi(1535)$ is completely analogous to the one of the $\Delta(1232)$, since, according to the PDG \cite{pdg} it couples to the $\pi\Xi$ channel with a branching ratio of $100\%$. Hence, the coupling $g_{\,\Xi^{*},\pi\Xi}$ is simply given by Eq. \eqref{eq:cophen}, doing the substitutions $M_{\Delta}\rightarrow M_{\Xi^{*}}$ and $M_{N}\rightarrow M_{\Xi}$ .

The case of the $\Omega^{-}$ is different since this resonance is stable to strong decays. This means that the on shell amplitude $\Gamma_{on}$ is zero and this prevents us from evaluating the coupling of the resonance to the $\bar{K}\Xi$ channel using Eq. \eqref{eq:cophen}. However, from $SU(3)$ symmetry considerations we can relate the $g_{\Omega^{-},\bar{K}\Xi}$ coupling to $g_{\Delta,\pi N}$, since their ratios are simply ratios of Clebsch-Gordon coefficients.

We find that
\begin{equation}
g_{\Omega^{-},\bar{K}\Xi^{0}}^2=2g_{\Delta,\pi N}^2\ .
\end{equation}

The amplitude in relativistic form is again given by Eq. \eqref{eq:phenampl} and, in the case of the $\Sigma(1385)$ and $\Xi(1535)$, it is extrapolated to the second Riemann sheet in order to evaluate the pole and the new couplings. Since, as already said, the $\Omega^{-}$ does not decay through strong interaction, the pole of the amplitude is found on the real axis, with no need to go to the second Riemann sheet. It is then possible to apply the sum-rule, evaluating the derivative of the $G$ function in the position of the pole. To do it we use a cutoff of the same order of magnitude of the one found doing the best fit for the $\Delta(1232)$, $q_{max}\simeq 450\ MeV$.  The results obtained for the three resonances are shown in Table \ref{tab:decuplet}. We also show the cutoff dependence of $1-Z$, analogous to Table \ref{tab:z}, in \ref{tab:decupletcutoff}.

\begin{table}[ tp ]%
\begin{tabular}{c||c|c|c|c|c}
\hline %
& Channel & $\sqrt{s_0}\ [MeV]$ & $g\ [MeV]^{-1}$ & $-g^2\frac{\partial G^{II}}{\partial E}$ & $1-Z$ \\\toprule %
$\Sigma(1385)$ & \begin{tabular}{c}$\pi\Lambda$ \\ $\pi\Sigma$\end{tabular} &  \begin{tabular}{@{}c@{}}$1380.36+i17.29$ \\ $1377.35+i16.02$\end{tabular} & \begin{tabular}{@{}c@{}}$(5.11+i0.60)\cdot 10^{-3}$ \\ $(3.63+i0.81)\cdot 10^{-3}$\end{tabular} & \begin{tabular}{@{}c@{}}$(0.16-i0.18)$ \\ $(9.62-i1.16)\cdot 10^{-2}$\end{tabular} & \begin{tabular}{@{}c@{}}$0.16$ \\ $0.10$\end{tabular} \\\hline
 $\Xi(1535)$ & $\pi\Xi$ & $1532.92+i4.68$ & $(4.36+i0.23)\cdot 10^{-3}$ & $0.11-0.09$ & $0.11$ \\\hline
 $\Omega^{-}$ & $\bar{K}\Xi$ & $1672.45$ & $(1.56+i0.37)\cdot 10^{-2}$  & $0.26$ & $0.26$\\\hline
\end{tabular}
\caption{Values of poles, couplings,  $-g^2\frac{dG^{II}}{d\sqrt{s}}$ and $1-Z$ for the three baryons of the decuplet $J^P=\frac{3}{2}^{+}$, $\Sigma(1385)$, $\Xi(1535)$ and $\Omega^{-}$, for a cutoff $q_{max}=450\ MeV$.}
\label{tab:decuplet}\centering %
\end{table}

\begin{table}[ tp ]%
\begin{tabular}{c||c|c|c|c}
\hline %
& Channel & $q_{max}$ [GeV] & $-g^2\frac{\partial G^{II}}{\partial E}$ &$1-Z$   \\\toprule %
$\Sigma(1385)$ & $\pi\Lambda$ & \begin{tabular}{@{}c@{}} $0.4$\\$0.5$\\ $0.6$\end{tabular}  & \begin{tabular}{@{}c@{}}$(0.13-i0.19)$ \\ $(0.19-i0.17)$\\ $(0.24-i0.16)$\end{tabular} & \begin{tabular}{@{}c@{}} $0.13$\\$0.19$\\ $0.24$\end{tabular}   \\\hline
$\Sigma(1385)$ & $\pi\Sigma$ & \begin{tabular}{@{}c@{}} $0.4$\\$0.5$\\ $0.6$\end{tabular}  & \begin{tabular}{@{}c@{}}$(8.71-i1.17)\cdot 10^{-2}$ \\ $(0.10-i6.42\cdot 10^{-3})$\\ $(0.12-i3.37\cdot 10^{-3})$\end{tabular} & \begin{tabular}{@{}c@{}} $0.09$\\$0.10$\\ $0.12$\end{tabular}   \\\hline
$\Xi(1535)$ & $\pi\Xi$ & \begin{tabular}{@{}c@{}} $0.4$\\$0.5$\\ $0.6$\end{tabular}  & \begin{tabular}{@{}c@{}}$(0.09-i0.09)$ \\ $(0.12-i0.09)$\\ $(0.15-i0.09)$\end{tabular} & \begin{tabular}{@{}c@{}} $0.09$\\$0.12$\\ $0.15$\end{tabular}   \\\hline
$\Omega^-$ & $\bar{K}\Xi$ & \begin{tabular}{@{}c@{}} $0.4$\\$0.5$\\ $0.6$\end{tabular}  & \begin{tabular}{@{}c@{}}$0.18$ \\ $0.34$\\ $0.53$\end{tabular} & \begin{tabular}{@{}c@{}} $0.18$\\$0.34$\\ $0.53$\end{tabular}   \\\hline
\end{tabular}
\caption{Values of $-g^2\frac{dG^{II}}{d\sqrt{s}}$ and  $1-Z$ for different cutoffs $q_{max}$ for the three baryons of the decuplet $J^P=\frac{3}{2}^{+}$, $\Sigma(1385)$, $\Xi(1535)$ and $\Omega^{-}$.}
\label{tab:decupletcutoff}\centering %
\end{table}

\section{Interpretation of the sum-rule for resonances}
\label{five}
As we could see, we have obtained values of $-g^2\frac{dG^{II}}{d\sqrt{s}}$ which are complex, and, thus, cannot literally be interpreted as a probability. In this Section we clarify the meaning of the sum-rule in Eq. \eqref{eq:srreal} and of the value of $1-Z$ obtained.

Before we give a general formulation of the sum-rule for complex energies based on the results of \cite{gamermann, yamagata, aceti}, let us visualize it in a particular case with two channels, one of them closed and the other one open. Let us also assume, for simplicity, that the interaction in the closed channel is strong and attractive and let us neglect the diagonal interaction in the open channel (the results are the same without this restriction, only the formulation is a little longer). Thus, we have a potential like in Eq. \eqref{eq:pot} but now
\begin{equation}
\label{eq:pot2x2}
v=
\begin{pmatrix}
v_{11} & v_{12} \\
v_{12} & 0 \end{pmatrix}\ .
\end{equation}  

The results that we get are  general, and including $v_{22}$ is straightforward but does not add to the discussion.  We shall also assume for simplicity that $|v_{12}|\ll|v_{11}|$ only to relate  the imaginary part of the pole position to the width. 

The $t$ matrix is given by Eq. \eqref{eq:t}, and we find
\begin{equation}
\label{eq:t2x2}
t=
\begin{pmatrix}
v_{11}+v_{12}^2G_{2} & v_{12} \\
v_{12} & v_{12}^2 G_{1}
\end{pmatrix}
\cdot \frac{1}{1-v_{11}G_{1}-v_{12}^2G_{1}G_{2}}\ .
\end{equation}

Let us now assume that we have a pole in the bound region of channel $1$ and open region of channel $2$. Then, the denominator of $t$ in Eq. \eqref{eq:t2x2} will be zero
\begin{equation}
\label{eq:det}
1-v_{11}G_{1}-v_{12}^2G_{1}G_{2}=0\ ,
\end{equation}
but $G_2$ is complex in the first Riemann sheet with
\begin{equation}
\label{eq:imclass}
Im G_{2}^{I}=-i4\pi^2\mu_{2}\,k^3_2\ 
\end{equation}
in the non-relativistic formulation, and
\begin{equation}
\label{eq:imrel}
Im G_{2}^{I}=-i\frac{1}{4\pi}\frac{M_N}{\sqrt{s}}\,k^3_2\ 
\end{equation} 
in the relativistic one of Section \ref{delta}, with $k_i=\sqrt{2\mu_i(E-m_{1i}-m_{2i})}$ or $k_i=\frac{\lambda^{1/2}(s,m_{1i}^2,m_{2i}^2)}{2\sqrt{s}}$ respectively, for $i=2$.

Let us assume that the attractive $v_{11}$ interaction is strong enough to produce a bound state in channel $1$ with energy $E_1$, when only this channel is considered. Then, we would have 
\begin{equation}
\label{eq:pole1}
1-v_{11}G_1(E_1)=0\ .
\end{equation}
The addition of the interaction $v_{12}$ will change this energy and Eq. \eqref{eq:det} can be rewritten, taking  Eq. \eqref{eq:pole1} into account, as (assume $v_{ij}$ independent of energy)
\begin{equation}
\label{eq:pole2}
-v_{11}\frac{\partial G_1}{\partial E}(E_R-E_1)-v_{12}^2G_1G_2=0\ ,
\end{equation}
where $E_R$ will be the new energy of the system.

Since $v_{11}<0$ and $\frac{\partial G_{1}}{\partial E}<0$ in the bound region
\begin{equation}
\label{eq:ener}
E_R-E_1=-\alpha\, v_{12}^2G_1G_2\ ,\ \ \ \ \ \ \ \ \ \ \ \ \ \alpha>0\ ,\ \ \ G_1<0\ .
\end{equation}

The complex value of $G_2$, see Eqs. \eqref{eq:imclass} and \eqref{eq:imrel}, was obtained for an energy $E+i\epsilon$. We gradually continue along the complex plane making the $i\epsilon$ finite, $i\frac{\Gamma}{2}$, and Eq. \eqref{eq:ener} gives
\begin{eqnarray}
\label{eq:ener2}
\tilde{E}_R+i\frac{\Gamma}{2}&=&-\alpha\,v_{12}^2G_1G_2\ , \\
\frac{\Gamma}{2}&\simeq & -\alpha\,v_{12}^2 G_1ImG_2\ ,
\end{eqnarray}
which is impossible to fulfill in the first Riemann sheet since  $G_1<0$, $\alpha>0$ and $ImG_2^{I}$, given by Eqs. \eqref{eq:imclass}-\eqref{eq:imrel}, is negative. This gives us a perspective of why one has to go to the second Riemann sheet, where $k_2\rightarrow -k_2$ in $G_2$, in which case one finds a solution, with $\tilde{E}_R=E_1-m_{1i}-m_{2i}$ ($i=2$) and
\begin{equation}
\label{eq:width}
\Gamma=2 \frac{v_{12}^2G_1}{-v_{11}\frac{\partial G_1}{\partial E}}ImG_{2}^{II}\ .
\end{equation}

Next, let us calculate the couplings $g_i$, where $g_ig_j$ is the residue of the $t_{ij}$ matrix element at the pole. Applying l'H\^{o}pital rule, we have
\begin{equation}
\label{eq:couplings}
\begin{split}
&g_1^2=\lim(E-E_R)t_{11}=\frac{v_{11}+v_{12}^2G_2}{-v_{11}\frac{\partial G_1}{\partial E}-v_{12}^2\frac{\partial G_1}{\partial E}G_2-v_{12}^2\frac{\partial G_2}{\partial E} G_1}\ , \\
&g_2^2=\lim(E-E_R)t_{22}=\frac{v_{12}^2G_1}{-v_{11}\frac{\partial G_1}{\partial E}-v_{12}^2\frac{\partial G_1}{\partial E}G_2-v_{12}^2\frac{\partial G_2}{\partial E} G_1}\ .
\end{split}
\end{equation}

Let us now see that the sum-rule of Eq. \eqref{eq:srreal} is exactly fulfilled, since we have
\begin{equation}
\label{eq:sumrulecomplex}
g_1^2\frac{\partial G_1}{\partial E}+g_2^2\frac{\partial G_2}{\partial E}=\frac{(v_{11}+v_{12}^2G_2)\frac{\partial G_1}{\partial E}+v_{12}^2G_1\frac{\partial G_2}{\partial E}}{-v_{11}\frac{\partial G_1}{\partial E}-v_{12}^2\frac{\partial G_1}{\partial E}G_2-v_{12}^2\frac{\partial G_2}{\partial E} G_1}=-1\ .
\end{equation}
However, this occurs only at the complex pole $\tilde{E}_R+i\frac{\Gamma}{2}$ using $G_2^{II}$, since we have made use of the fact that the denominator in $g_1^2$ and $g_2^2$ of Eqs. \eqref{eq:couplings} vanishes for $E=\tilde{E}_R+i\frac{\Gamma}{2}$ to apply l'H\^{o}pital rule, which only occurs in the second Riemann sheet.

Note that the sum-rule has appeared with the definition of the couplings of Eq. \eqref{eq:coupling}. The explicit form obtained for the couplings in Eqs. \eqref{eq:couplings} shows clearly that they are complex, since both $G_1$ and $G_2$ are now complex.

Now that we have obtained the couplings, let us rewrite $\Gamma$ of Eq. \eqref{eq:width}, derived assuming $|v_{12}|\ll|v_{11}|$ and neglecting again $v_{12}$ versus $v_{11}$, as
\begin{equation}
g_2^2\simeq \frac{v_{12}^2G_1}{-v_{11}\frac{\partial G_1}{\partial E}}\ ,
\end{equation}
from which follows
\begin{equation}
\label{eq:width2}
\Gamma=2 g_2^2\frac{M}{4\pi\sqrt{s}}p^3\ ,
\end{equation}
where we have used the relativistic formula for $ImG_2$ of Eq. \eqref{eq:imrel} and Eq. \eqref{eq:GII}. As we can see, we reproduce the formula for the width given by Eq. \eqref{eq:cophen}.

Now we want to interpret the meaning of the sum-rule. Eq. \eqref{eq:sumrulecomplex} is a generalization to complex energies of the sum-rule obtained in Eq. (119) of \cite{gamermann} and Eq. (101) of \cite{aceti} for real energies. There it was interpreted as a consequence of the sum of probabilities of each channel to be unity. For complex values of the energies this interpretation is not possible and this is related to the fact that the eigenstates of a complex Hamiltonian are not generally orthogonal\footnote{Although our Hamiltonian was given in terms of $v_{ij}$ in coupled channels, only for formal purposes one could think of a complex Hamiltonian whose eigenvalues would be these complex energies. }. 

Formally the problem is solved using, in this case, a biorthogonal basis. Indeed, let $\lambda_n$ be a complex eigenvalue of $H$ and $|\lambda_n\rangle$ the corresponding eigenvector. It satisfies
\begin{equation}
det(H-\lambda_n I)=0\ .
\end{equation} 
Then 
\begin{equation}
det(H^{\dagger}-\lambda_n^{*} I)=0\ ,
\end{equation} 
which means that $\lambda_n^{*}$ is an eigenvalue of $H^{\dagger}$. Let now $|\bar{\lambda}_n\rangle$ be the eigenvector of $H^{\dagger}$ associated to $\lambda_n^{*}$. The eigenvectors $|\lambda_n\rangle$ and $|\bar{\lambda}_n\rangle$ are not equal, but we can see that 
\begin{equation}
\langle\bar{\lambda}_n|H|\lambda_m\rangle=\lambda_m\langle\bar{\lambda}_n|\lambda_m\rangle=\lambda_n\langle\bar{\lambda}_n|\lambda_m\rangle\ ,
\end{equation} 
where to get the last term we have applied $H$ as $H^{\dagger}$ to the bra state. Thus
\begin{equation}
(\lambda_n-\lambda_m)\langle\bar{\lambda}_n|\lambda_m\rangle=0\ ,
\end{equation}
which means that $|\lambda_m\rangle$ and $|\bar{\lambda}_n\rangle$ are orthogonal for $n\neq m$. For the case of $n=m$, $\langle\bar{\lambda}_n|\lambda_n\rangle\neq 0$ and we can choose a normalization and a phase for $|\bar{\lambda}_n\rangle$ and $|\lambda_n\rangle$ such that $\langle\bar{\lambda}_n|\lambda_n\rangle= 1$.

The resolution of the identity is now given by $\sum_n |\lambda_n\rangle\langle\bar{\lambda}_n|$. Furthermore, if we have a symmetric but not hermitian Hamiltonian, as it is our case, then it is trivial to see that $|\bar{\lambda}_n\rangle=|\lambda^{*}_n\rangle$ for its wave function. 

Then, the relationship 
\begin{equation}
\langle\Psi_i|\Psi_i\rangle=\sum_i \int d^3p|\Psi_i(p)|^2=1
\end{equation}
used to derive the sum-rule in \cite{gamermann,aceti}, must be substituted by 
\begin{equation}
\label{eq:integral}
\langle\bar{\Psi}_i|\Psi_i\rangle=\sum_i \int d^3p\,(\bar{\Psi}_i(p))^{*}\Psi_i(p)=\int d^3p\,\Psi^2_i(p)=1\ .
\end{equation}
Hence, for complex values, the modulus squared of the wave function has to be substituted by its square. The integral of Eq. \eqref{eq:integral} depends on the prescription used for the phase of $\Psi_i$. Below we show that with the standard phase convention used in \cite{aceti}, Eq. \eqref{eq:integral} is fulfilled.

Recalling that the wave function for us is given by (omitting the spherical harmonics) \cite{aceti}\footnote{This wave function is for a decaying channel of the resonance (it does not have the $|\Phi\rangle$ term in the wave function in Eq. \eqref{eq:2}). One can assume that in the formalism of \cite{yamagata, aceti} (see Eqs. (46) and (47) in \cite{yamagata}) the asymptotic scattering state used to create the resonance couples extremely weakly to it, such that one only has to worry for the sum-rule about the bound state and the relevant decaying states which have the wave function of Eq. \eqref{eq:wf}.}
\begin{equation}
\label{eq:wf}
\Psi_i(p)=g_i\frac{\Theta(\Lambda-|\vec{p}\,|)p}{E-m_{1i}-m_{2i}-p^2/2\mu_i}\ ,
\end{equation}
we can write
\begin{equation}
\label{eq:wf2}
\int d^3p\,(\Psi(p))^2=g_i^2\int_{|\vec{p}\,|<\Lambda}d^3p\frac{p^2}{(E-m_{1i}-m_{2i}-p^2/2\mu_i)^2}=-g_i^2\frac{\partial G_i^{II}}{\partial E}\ ,
\end{equation}
but we saw in Eq. \eqref{eq:sumrulecomplex} that
\begin{equation}
\sum_i g_i^2\frac{\partial G_i^{II}}{\partial E}=-1\ ,
\end{equation}
and, hence, we conclude that
\begin{equation}
\int d^3p\,\Psi^2_i(p)=1\ ,
\end{equation}
with the phase and normalization chosen for the wave function in Eq. \eqref{eq:wf}.

Note that for the case of bound states we can use the same formulation and the prescription taken for the phase is the one where the wave function is real. In general $g_i$ can be complex and $\Psi_i(p)$ will be complex, but the prescription for the interpretation given is to take the phase convention with the wave function in momentum space given by Eq. \eqref{eq:wf}.

This clarifies the meaning of the sum-rule. It is the demanded extrapolation to complex energies of the sum of probabilities equal unity for real energies. The modulus square of the wave function is substituted by the square of the wave function with a given prescription for the phase, which in the case of bound states would be having the wave function real. Thus we should interpret $-g_i^2\frac{\partial G_i^{II}}{\partial E}$ as the extrapolation of a probability into the complex plane, but it is not a probability. Yet, once we have interpreted it as the integrated strength of the wave function squared, we still can think of it as a magnitude providing the weight, or relevance of one given channel in the wave function of a state.

As we can see, the integral $\int d^3p\,(\Psi(p))^2$, given in terms of the coupling $g_i$ and $\frac{\partial G_i^{II}}{\partial E}$, is a finite but complex quantity.

Since the two terms in Eq. \eqref{eq:sumrulecomplex} will now be complex, the sum of the imaginary parts will vanish and the sum of real parts will be equal to $-1$. Thus we have
\begin{equation}
Re(g_1^2\frac{\partial G_1^{II}}{\partial E})+Re(g_2^2\frac{\partial G_2^{II}}{\partial E})=-1
\end{equation}
and the sum-rule is fulfilled for the real part of the squared of the wave functions. 

The evaluation of the integral of $(\Psi_i(p))^2$ is most easily done in momentum space and concretely in terms of $G^{II}$. Yet, one would like to have a feeling of what happens in terms of wave functions in coordinate space, even if the integration of $(\Psi_i(r))^2$ in coordinate space requires extra work and is not convenient. We calculate the wave function in coordinate space in Appendix \ref{appendix} and we recall only the basic results that we use here for qualitative purposes. 

For $r\rightarrow \infty$ one obtains for the open channel, in the non relativistic formulation of Section \ref{form} and in the first Riemann sheet
\begin{equation}
\label{eq:wfcoord}
\Psi_2(r)\sim\frac{e^{ikr}}{r}\ ,\ \ \ \ \ \ \ \ k=\sqrt{2\mu_2(\tilde{E}_R+i\frac{\Gamma}{2})}\ .
\end{equation}
Defining $k_R=\sqrt{2\mu_2\tilde{E}_R}$ and $k_I=\sqrt{2\mu_2\tilde{E}_R}\frac{\Gamma}{4\tilde{E}_R}$, we can write
\begin{equation}
\label{eq:wfcoord2}
\Psi_2(r)\sim\frac{1}{r}e^{ik_Rr}e^{-k_Ir}\ .
\end{equation}
In the second Riemann sheet, we would substitute $k$ by $-k$ and then
\begin{equation}
\label{eq:wfcoord2}
\Psi_2^{II}(r)\sim\frac{1}{r}e^{-ik_Rr}e^{k_Ir}\ .
\end{equation}
Hence the wave function in coordinate space in the second Riemann sheet would even blow up, such that a probability would be infinite. This is actually also the case even if $k_I=0$. Thus the concept of probability is not useful once we have open channels.

Yet,
\begin{equation}
\label{eq:wfcoord3}
(\Psi_2^{II}(r))^2\sim\frac{1}{r^2}e^{-2ik_Rr}e^{2k_Ir}\ ,
\end{equation}
and it has an oscillatory behavior that makes the integral for large values of $r$ vanish in the sense of a distribution, like $\int d^3r\, e^{i\vec{p}\,\vec{r}}$ for $p\neq 0$. Of course, the finiteness of the integral is better seen integrating in the space of momenta, as we have seen.


\section{Interpretation of the sum-rule for energy dependent potentials}

We have been using a potential that has a CDD pole as Eq. \eqref{eq:pot1} and have interpreted $-g^2\frac{\partial G}{\partial E}$ as the probability, in the case of bound states, of having a certain state. The sum-rule of Eq. \eqref{eq:sumrule} holds for states that are generated in coupled channels with potentials which are independent of the energy. Yet, in the case on an energy dependent potential, like the one of Eq. \eqref{eq:pot1}, we still would associate $-g^2\frac{\partial G}{\partial E}$ to the possibility of finding channel $1$ in a certain state (for bound states). This requires a justification. One can in principle go back to the work done in \cite{gamermann} and rederive the formulas with an energy dependent potential. However, this meets with serious problems, because the eigenstates of the Hamiltonian are now not orthogonal and $\sum|\alpha\rangle\langle\alpha|$ is not the resolution of the identity. These problems and possible solutions are studied in \cite{sazdjian, mares}.

A suggestion to interpret the results of the sum-rule in the case of an energy dependent potential is given in  \cite{hyodorep} in Section 4.2. For a state generated  with an energy dependent potential in coupled channels, one has a sum-rule
\begin{equation}
\label{eq:srenergy}
-\sum_{i,j}g_ig_j\left[\frac{\partial G_i^{II}(E)}{\partial E}\delta_{ij}+G_i^{II}(E)\frac{\partial V_{ij}(E)}{\partial E}G_j^{II}(E)\right]_{E\rightarrow E_R}=1\ ,
\end{equation}
with  $V_{ij}(E)$ the interaction kernel, and then the compositeness (the probability of the state to be in either of the channels considered, for the case of bound states) is defined as
\begin{equation}
X=-\sum_i g_i^2\left[\frac{\partial G_i^{II}(E)}{\partial E}\right]_{E\rightarrow E_R}\ ,
\label{eq:compo}
\end{equation}
while the elementariness (the part of the state that does not belong to the considered channels) is then
\begin{equation}
Z=-\sum_{ij} \left[g_i G^{II}_i(E)\frac{\partial V_{ij}(E)}{\partial E}G_j^{II}(E)g_j\right]_{E\rightarrow E_R}\ .
\label{eq:element}
\end{equation}

We take advantage here to justify this in the case that we have discussed above with two channels, and we study the problem with a single channel and an effective energy dependent single channel potential. The effective potential method is also discussed in \cite{hyodorep} in Section 3.2, using the Feshbach projection method \cite{feshbach}. Here we follow a different approach. 

The idea is the following: we start with a two channel case with an energy independent potential that generates a certain bound state, and evaluate $T_{ij}$. Then we use just channel 1 with an effective potential, such that $T_{11}$ is the same in both approaches. As a consequence, and for bound states, $-g_1^2\frac{\partial G_1}{\partial E}$, which is the same in both approaches, gives the probability to find channel 1 in the state that we study, which is smaller than one. The difference from unity of this quantity, in our approach, is $Z$, which gives the probability that the state that we find is not in channel 1. This latter probability is related to $\frac{\partial V_{eff}}{\partial E}$ as we see below. 

Indeed, using the simplified case of Eq. \eqref{eq:pot2x2} in two channels we have
\begin{equation}
\label{eq:amplsimp}
T_{11}=\frac{v_{11}+v_{12}^2G_2}{1-(v_{11}+v_{12}^2G_2)G_1}\ ,
\end{equation}
while in one channel with $V_{eff}$, we will have
\begin{equation}
\label{eq:teff}
T_{eff}=\frac{V_{eff}}{1-V_{eff}G_1}\ .
\end{equation}
It is clear that taking
\begin{equation}
\label{eq:veff}
V_{eff}=v_{11}+v_{12}^2G_2
\end{equation}
 the two amplitudes $T_{11}$ and $T_{eff}$ are identical and the residue at the pole, $g_1^2$, will also be the same as $g_{eff}^2$.

On the other hand, we have from Eq. \eqref{eq:coupling}
\begin{equation}
\label{eq:geff}
g_{eff}^2=\lim \frac{(E-E_0)V_{eff}} {1-V_{eff}G_1}=\frac{V_{eff}}{-\frac{\partial V_{eff}}{\partial E}G_1-V_{eff}\frac{\partial G_1}{\partial E}} \ ,
\end{equation}
which, using the pole condition $1-V_{eff}G_1=0$, can be rewritten as
\begin{equation}
\label{eq:geff2}
g_{eff}^2=\frac{1}{-G_1^2\frac{\partial V_{eff}}{\partial E}-\frac{\partial G_1}{\partial E}}\ .
\end{equation}
Hence,
\begin{equation}
\label{eq:geff3}
-g_{eff}^2 G_1^2\frac{\partial V_{eff}}{\partial E}-g_{eff}^2\frac{\partial G_1}{\partial E}=1\ .
\end{equation}

Since 
\begin{equation}
\label{eq:x}
X\equiv -g_{eff}^2\frac{\partial G_1}{\partial E}=-g_{1}^2\frac{\partial G_1}{\partial E}
\end{equation}
is the probability to find the state in  channel 1 (for bound states), then
\begin{equation}
\label{eq:zeff}
-g_{eff}^2G_1^2\frac{\partial V_{eff}}{\partial E}\equiv Z
\end{equation}
gives the probability to find the state somewhere else (originally channel 2). This is the result of \cite{hyodorep} which we wrote in Eq. \eqref{eq:element}.

Note that to study the possibility to have a genuine (non $\pi N$) state in the resonance that we study, we have used a CDD pole term in the potential. We can use this also to account for missing channels. Coming back to our example, if we have (we change a bit the notation for convenience)
\begin{equation}
\label{eq:veff2}
V_{eff}=a+\frac{b}{E-E_R}
\end{equation} 
we should equate it to the potential of Eq. \eqref{eq:veff}, which is not possible in all the range of energies. But the minimum requirement is that they are the same at the pole and give the same residue, for which it suffices to equate the two values of $\frac{\partial V_{eff}}{\partial E}$. Thus,
\begin{equation}
\begin{split}
&a+\frac{b}{E_0-E_R}=v_{11}+v_{12}^2G_2(E_0)\ ,\\
&-\frac{b}{(E_0-E_R)^2}=v_{12}^2\left[\frac{ \partial G_2}{\partial E}\right]_{E=E_0}\ .
\end{split}
\end{equation}
This is always possible and shall leave us still one parameter to make a fit for an optimal agreement of the two expressions in a certain range of energies around the pole.

Finally, let us make a small remark in the sense that, indeed, the use of the CDD pole is a suited way to take into account the genuine states in a problem. For this, we take just the CDD pole term in $V_{eff}$ with a small coupling $b$ to channel 1. We then should expect to get $Z\simeq 1$, which is just the case. Indeed, 
\begin{equation}
\begin{split}
\label{eq:teffeff}
&T_{eff}=\frac{1}{\frac{E-E_R}{b}-G_1}\ ,\\
&g_{eff}^2=\frac{1}{\frac{1}{b}-\frac{\partial G_1}{\partial E}}\ ,
\end{split}
\end{equation}
and when $b\rightarrow 0$ then $g_{eff}^2\simeq b$. Hence
\begin{equation}
\begin{split}
&X=-g_{eff}^2\frac{\partial G_1}{\partial E}\rightarrow 0\ ,\\
&Z=-g_{eff}^2G_1^2\frac{\partial V_{eff}}{\partial E}\rightarrow -bG_1^2(E_0)\frac{-b}{(E_0-E)^2}=\left(\frac{b}{E_0-E_R}G_1·\right)^2=1\ ,
\end{split}
\end{equation}
the last equation holding because of the pole in the denominator of $T_{eff}$, Eq. \eqref{eq:teffeff}.

\section{Discussion and Conclusions}
\label{concl}
We have applied the generalized compositeness condition to the decuplet of the $\Delta(1232)$ to see the weight of meson-baryon cloud and genuine (presumably three quark) components. It is interesting to see that we find the pole position for the $\Delta(1232)$, Eq. \eqref{eq:polebf}, in very good agreement with the PDG \cite{pdg} values.

We clarified here the meaning of the extension of the Weinberg sum-rule for the case of resonances and found that $-g^2\frac{\partial G_{II}}{\partial E}$ measures $\int d^3p\ \langle\vec{p}\ |\Psi\rangle^2$ and not  $\int d^3p\ |\langle\vec{p}\ |\Psi\rangle|^2$. We found that the integral of the real part of the square of the wave function is the natural quantity to provide a measure of the relevance of an open channel in the wave function, since the integral of the modulus squared diverges, even more in the second Riemann sheet. On  the other hand, $\int d^3p\ \langle\vec{p}\ |\Psi\rangle^2$ is finite and the sum of these quantities for the different coupled channels is unity, within a certain phase convention, as shown by the generalization of the Weinberg sum-rule. 
As to the weight of the $\pi N$ component in the $\Delta(1232)$ wave function, we find values which are relatively high, of the order of 60 \%. This number could sound a bit large when one thinks of the $\Delta(1232)$ as just a spin flip on the quark spins of the nucleon. Yet, the result is less surprising when one recalls that from Drell Yan and deep inelastic scattering one induces a probability of about 34 \% for the $\pi N$ component in the nucleon \cite{peng1,peng2}. When one realizes this, then it also looks less surprising that, unlike the case of the $\rho$, where the analysis in terms of just the $\pi \pi$ component requires large counterterms beyond the lowest order contribution from the chiral Lagrangians \cite{ramonet,ollerpalomar}, in the case of the $\pi N$ scattering in the $\Delta(1232)$ region a description was possible with moderate size of the counterterms \cite{juan1,juan2}.

We extended the compositeness test to the other members of the decuplet and found a decreasing size of the meson-baryon components when we go to the $\Sigma(1385)$ and $\Xi(1535)$, indicating that the higher energy members of the decuplet are better represented by a genuine (in principle three quark) component. For the $\Sigma(1385)$ and $\Xi(1535)$ there are also bound components of $\bar{K} N$ and $\bar{K} \Lambda$, $\bar{K} \Sigma$, respectively, which we estimate small compared to the open ones in the limited space allowed due to the decay into the open components. In the case of the $\Omega^-$, where only the bound component $\bar{K} \Xi$ is present, we estimate the weight of the meson-baryon component to be small, of the order of 25 \%.

The large pion nucleon cloud in the $\Delta(1232)$ indicates that realistic calculations of its properties should take this cloud into account. Even before the present test was done to estimate the weight of $\pi N$ component in the $\Delta(1232)$ wave function, the importance of the meson cloud has been often advocated and one example of it can be seen in the early works on the cloudy bag model \cite{cbm} or chiral quark model \cite{chiqm}. The work presented here offers a new perspective on this interesting subject and the possibility to become more quantitative than in early works. 

We have also taken advantage to find an interpretation of the extension of the Weinberg sum-rule for complex values of the energy. We found  that the concept of probability is then changed  to the squared of the wave function, within a certain phase convention, which, upon integration, leads to finite values that we present as a measure of the weight of a channel in the wave function, while the modulus squared of the wave function is divergent for open channels.

We have also given an interpretation of the terms of the sum-rule for the case of an energy dependent potential. In the case that we have a complete set of coupled channels that generates a certain bound state, we can truncate the space and  define an energy dependent potential in a space of lower dimension. The sum-rule is now rewritten and a physical interpretation is given to the different terms. The probability $Z$ that the state overlaps with the eliminated part of the space is related to the derivative of the potential with respect to the energy.
\appendix
\section{Wave functions in coordinate space}
\label{appendix}
The wave function in momentum space is given by (let us take also a spherical harmonic $Y_{10}(\hat{p})$ for simplicity)
\begin{eqnarray}
\label{eq:a}
\Psi(\vec{p}\,)=g\frac{\theta(\Lambda-p)\,p}{E-p^2/2\mu}Y_{10}(\hat{p})\equiv \tilde{\Psi}(\vec{p}\,)Y_{10}(\hat{p})\ .
\end{eqnarray}
In coordinate space we can write
\begin{equation}
\label{eq:a1}
\Psi(\vec{r}\,)=\int \frac{d^3p}{(2\pi)^{3/2}}e^{i\vec{p}\,\vec{r}}\Psi(\vec{p})\ .
\end{equation}
Integrating over the coordinate space we have
\begin{equation}
\label{eq:a2}
\begin{split}
\int d^3r\,(\Psi(\vec{r}))^2&=\int d^3r\,\int \frac{d^3p}{(2\pi)^{3/2}}e^{i\vec{p}\,\vec{r}}\Psi(\vec{p})\int \frac{d^3p\,'}{(2\pi)^{3/2}}e^{i\vec{p}\,'\vec{r}}\Psi(\vec{p}\,')\\ &=\int d^3p\,\int d^3p\,'\,\Psi(\vec{p})\Psi(\vec{p}\,')\delta(\vec{p}+\vec{p}\,')\ .
\end{split}
\end{equation}
Then, since
\begin{equation}
\label{eq:a3}
\Psi(\vec{p}\,')=\Psi(-\vec{p}\,)=(-1)^l\Psi(\vec{p}\,)\ 
\end{equation}
with $l=1$, we get
\begin{equation}
\label{eq:a4}
\int d^3r\,(\Psi(\vec{r}\,))^2=-\int d^3p\,(\Psi(\vec{p}\,))^2\ .
\end{equation}

In Eq. (39) of \cite{aceti} we found that, in this case
\begin{equation}
\label{eq:a5}
\Psi(\vec{r}\,)=\tilde{\Psi}(\vec{r}\,)Y_{10}(\hat{r}\,)
\end{equation}
and
\begin{equation}
\label{eq:a6}
\tilde{\Psi}(\vec{r}\,)=g\int_{p<\Lambda} \frac{d^3p}{(2\pi)^{3/2}}\,i\,j_{1}(pr)\frac{p}{E-p^2/2\mu}\ .
\end{equation}

If we remove the factor $i$ in $\tilde{\Psi}(\vec{r}\,)$ and call
\begin{equation}
\label{eq:a7}
\bar{\Psi}(\vec{r}\,)=g\int_{p<\Lambda} \frac{d^3p}{(2\pi)^{3/2}}\,j_{1}(pr)\frac{p}{E-p^2/2\mu}\ ,
\end{equation} 
then we see that
\begin{equation}
\label{eq:a8}
\int d^3p\,(\tilde{\Psi}(\vec{p}\,))^2=-\int d^3p\,(\bar{\Psi}(\vec{p}\,))^2\ .
\end{equation}
By using explicitly that
\begin{equation}
\label{eq:a9}
j_{1}(x)=\frac{\sin{x}}{x^2}-\frac{\cos{x}}{x}
\end{equation}
and using the symmetry of the integral in Eq. \eqref{eq:a7}, we can write
\begin{equation}
\label{eq:a10}
\bar{\Psi}(\vec{r}\,)=-2\mu g\frac{1}{2i}\frac{4\pi}{(2\pi)^{3/2}}\frac{1}{r^2}\int_{-\Lambda}^{\Lambda}dp\, \frac{p}{p^2-2\mu E}e^{ipr}+2\mu g\frac{1}{2}\frac{4\pi}{(2\pi)^{3/2}}\frac{1}{r}\int_{-\Lambda}^{\Lambda}dp\, \frac{p^2}{p^2-2\mu E}e^{ipr}\ .
\end{equation}

\begin{figure}[h!]
\includegraphics[width=9cm,height=6cm]{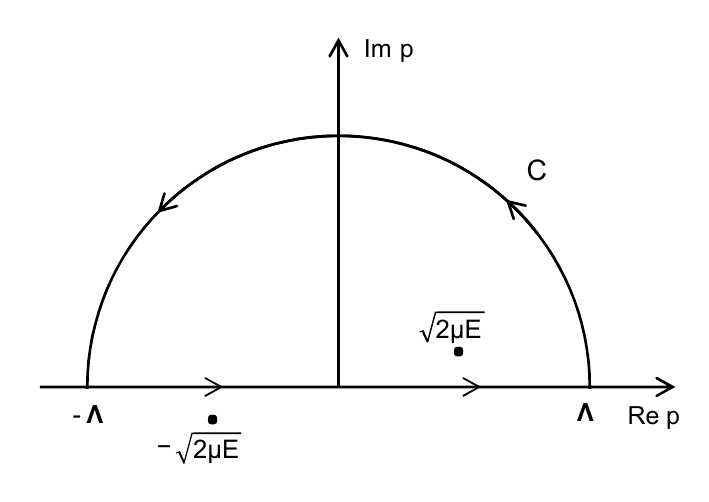}
\caption{Integration path in the complex $p$ plane for the wave function.}
\label{fig:int}
\end{figure}
We can perform the integration in $p$ by integrating over the circuit in the complex plane of Fig. \ref{fig:int}, and thus
\begin{equation}
\label{eq:a11}
\int_{-\Lambda}^{\Lambda}dp\,...=2\pi i\, Res(p=\sqrt{2\mu E})-\int_{\mathcal{C}}dp\, ...\ .
\end{equation}
The circuit picks up the pole at $p=\sqrt{2\mu E}$ and for the first Riemann sheet we find the result
\begin{equation}
\label{eq:a12}
\begin{split}
\bar{\Psi}(\vec{r}\,)=&-\mu g\frac{4\pi^2}{(2\pi)^{3/2}}\frac{1}{r^2}e^{i\sqrt{2\mu E}\,r}+\mu g i\frac{4\pi^2}{(2\pi)^{3/2}}\frac{1}{r}\sqrt{2\mu E}e^{i\sqrt{2\mu E}\,r}\\ &+\mu g\frac{4\pi}{(2\pi)^{3/2}}\frac{1}{r^2}\int_{0}^{\pi}d\theta\frac{\Lambda^2\,e^{2i\theta}}{\Lambda^2\,e^{2i\theta}-2\mu E}e^{i\Lambda r \cos\theta}e^{-\Lambda r \sin\theta}\\ &-\mu g\frac{4\pi}{(2\pi)^{3/2}}\frac{i}{r}\int_{0}^{\pi}d\theta\frac{\Lambda^3\,e^{3i\theta}}{\Lambda^2\,e^{2i\theta}-2\mu E}e^{i\Lambda r \cos\theta}e^{-\Lambda r \sin\theta}\ .
\end{split}
\end{equation}
In the second Riemann sheet we change $\sqrt{2\mu E}$ to $-\sqrt{2\mu E}$. As we can see, for large values of $r$ the integrals over the half circle in Fig. \ref{fig:int} are strongly suppressed by the factor $e^{-\Lambda r \sin\theta}$ ($\theta\in[0,\pi]$), which makes these integrals vanish when $r\rightarrow\infty$.

Then, the dominant term for $r\rightarrow\infty$ is given by
\begin{equation}
\label{eq:a13}
\bar{\Psi}^{II}(\vec{r}\,)\simeq -i\frac{1}{r}\sqrt{2\mu E}\,e^{-i\sqrt{2\mu E}\,r}\ ,
\end{equation}  
which has been used in the discussion in Section \ref{five}.


\section*{Acknowledgments}
We acknowledge the hospitality of the Kavli Institute for Theoretical Physics China. We also thank Miguel Albaladejo for a careful reading of the paper.
This work is partly supported by the Spanish Ministerio de Economia y Competitividad and European FEDER funds under the contract number FIS2011-28853-C02-01, the Generalitat Valenciana in the program Prometeo, 2009/090,  the Natural Science Foundation of China under the Grants Number 11375080 and 10975068, the Natural Science Foundation of Liaoning Scientific Committee (2013020091) and the Project of Knowledge Innovation Program (PKIP) of the Chinese Academy of Sciences under the Grants Number KJCX2.YW.W10.   F. Aceti thanks the Ministerio de Economia y Competitividad for the Beca de FPI. L.S. Geng acknowledges support  from the National Natural Science Foundation of China under Grant No. 11005007. We acknowledge the support of the European Community-Research Infrastructure Integrating Activity Study of Strongly Interacting Matter (acronym HadronPhysics3, Grant Agreement n. 283286) under the Seventh Framework Programme of EU.

\end{document}